\documentclass{svproc}
\usepackage{url}

\usepackage{graphicx}%
\usepackage{multirow}%
\usepackage{amsmath,amssymb,amsfonts}%
\usepackage{mathrsfs}%
\usepackage[mathscr]{eucal}
\usepackage[title]{appendix}%
\usepackage{xcolor}%
\usepackage{textcomp}%
\usepackage{manyfoot}%
\usepackage{booktabs}%
\usepackage[ruled,linesnumbered]{algorithm2e}
\usepackage[square,numbers]{natbib}
\usepackage{bbding}
\usepackage{verbatim}
\usepackage{comment}
\usepackage{booktabs,siunitx,ragged2e}
\newcolumntype{L}[1]{>{\RaggedRight\arraybackslash}p{#1}}

\begin{document}

\thispagestyle{empty}

\mainmatter              
\noindent This is a preprint of the paper:
\newline

\noindent\textit{Farzana Kabir, David Meg\'ias, and Krzysztof Cabaj.``RIOT-based smart metering system for privacy-preserving data aggregation using watermarking and encryption." Proceedings of the Computing Conference, June 2025, London (UK).} 

\setcounter{page}{0}

\pagebreak 
\title{RIOT-based smart metering system for privacy-preserving data aggregation using watermarking and  encryption}
\titlerunning{RIOT-based smart metering system}  

\author{Farzana Kabir\inst{1} \and David Megías \inst{1} \and Krzysztof Cabaj\inst{2}}
\authorrunning{Farzana Kabir et al.} 
%
\tocauthor{Farzana Kabir, David Megías, Cabaj Krzysztof}

\institute{Internet Interdisciplinary Institute (IN3), Universitat Oberta de Catalunya (UOC), CYBERCAT-Center for Cybersecurity Research of Catalonia, Rambla del Poblenou, 154-156, 08018 Barcelona, Spain\\
\and
Warsaw University of Technology, Plac Politechniki 1, 00-661 Warszawa, Poland}

\maketitle              

\abstract{The remarkable advancement of smart grid technology in the IoT sector has raised concerns over the privacy and security of the data collected and transferred in real-time. Smart meters generate detailed information about consumers' energy consumption patterns, increasing the risks of data breaches, identity theft, and other forms of cyber attacks. This study proposes a privacy-preserving data aggregation protocol that uses reversible watermarking and AES cryptography to ensure the security and privacy of the data. There are two versions of the protocol: one for low-frequency smart meters that uses LSB-shifting-based reversible watermarking (RLS) and another for high-frequency smart meters that uses difference expansion-based reversible watermarking (RDE). This enables the aggregation of smart meter data, maintaining confidentiality, integrity, and authenticity. The proposed protocol significantly enhances privacy-preserving measures for smart metering systems, conducting an experimental evaluation with real hardware implementation using Nucleo microcontroller boards and the RIOT operating system and comparing the results to existing security schemes.}

\keywords{Smart meter, Watermarking, Encryption, Privacy, RIOT}

\section{Introduction}\label{sec1}

Smart grids gather and analyze energy consumption data using IoT technology to enhance sustainability, efficiency, and reliability \cite{Orlando2022}. Smart metering systems are an integral part of the smart grid that collects, transmits, and evaluates consumers' energy consumption data in real-time using various means of communication, such as wireless sensor networks. These systems help energy service providers and consumers by providing a precise, reliable, and effective means of monitoring and controlling power consumption data \cite{Abdalzaher2022}. Due to the innovation of new and more complex features and functionalities installed into the smart metering system, security and privacy concerns are also multiplying and need to be seriously considered \cite{kua2023privacy}. Some security issues are related to user privacy, while others are associated with cyber-security due to the possibility of unauthorized access to smart meters and other entities in the system \cite{Xu2023, Bhola2022}. Hence, protecting the smart meters from these threats is now of the utmost importance for the researchers \cite{bavcnar2022security}.

\subsection{Motivation}\label{sec1.1}
A low-frequency smart meter collects energy consumption data every 1/2 or 1 hour and a high-frequency smart meter every 1 or 2 minutes \cite{kabir2024privacy}, both of which are enormous amounts of data to handle. Data aggregation is one of the leading concepts to protect users' fine-grained data in the smart metering system. Developing a robust data aggregation protocol that protects privacy, security, and the integrity of energy data is a challenging task \cite{kabir2021study}. Academics and industry people have been very interested in finding ways to increase the performance and efficiency of smart meter data while guaranteeing security, and numerous privacy-preserving solutions have been put forth \cite{wang2023privacy}. Cryptographic techniques like holomorphic encryption (HE) and data-hiding techniques like digital watermarking are mostly used in recent security solutions. With HE, calculations can be done directly on encrypted data (ciphertexts), and it offers a higher level of security. Nevertheless, a homomorphic cryptosystem's suitability for practical use depends on resource efficiency \cite{doan2023survey}. Because smart meters have limited processing power and resources, HE cannot be convenient due to complex computational tasks and large cipher texts. On the other hand, digital watermarking techniques use light computational operations to guarantee the authenticity and integrity of the data, but they cannot provide a high level of security alone \cite{kabir2023watercrypt}.

Modern operating systems dedicated to IoT-embedded devices, such as RIOT operating system, have built-in multiple cryptographic packages specifically optimized for running on ARM Cortex-M like STMicroelectronics (STM). \cite{rzepka2022performance}. These libraries are usually designed to be resource-efficient and optimized for the specific architecture of the microcontroller.

\subsection{Contribution}\label{sec1.2}
In this paper, we propose a privacy-preserving data aggregation solution for both high-frequency and low-frequency smart meters. To ensure the practicality of our solution, the protocol was implemented and tested using RIOT, an open-source operating system specifically designed for embedded IoT devices, and the ARM Cortex-M3 microcontroller (Nucleo-f207zg), a widely used hardware platform for IoT development. A key feature of RIOT is its built-in cryptographic support, which includes optimized implementations of encryption algorithms like AES (Advanced Encryption Standard). These features make RIOT an ideal choice for smart metering systems, where lightweight and secure data handling is crucial. The Nucleo board offers a balance of performance and energy efficiency, with enough computational power to handle complex data encryption tasks while maintaining energy efficiency. Our solution combines reversible watermarking and AES cryptography to achieve resource efficiency, lightweightness, and effectiveness.
The following are the main contributions of this research work:

\begin{itemize}

    \item A privacy-preserving protocol for low-frequency smart meters has been proposed using reversible watermarking based on LSB-shifting (RLS) and AES encryption.
    
    \item A privacy-preserving protocol for high-frequency smart meters has been proposed using reversible watermarking based on difference expansion (RDE) and AES encryption.
    
    \item A comparative analysis of the latest research works with the proposed schemes is offered to ensure the scheme ensures all the requirements with conceptual proofs, taking into account security and privacy needs and resistance to potential attacks.
    
    \item The proposed protocol is proven efficient and reliable by the experimental results obtained from a real hardware implementation using the Nucleo development board and the RIOT operating system, which results in minimal energy and time consumption.
    
\end{itemize}

The rest of the paper is organized as follows:
Section \ref{sec2} briefly reviews some of the recent studies related to privacy-preserving data aggregation in smart metering systems. Section \ref{sec3} describes the overall representation of the system model, and Section \ref{sec4} describes the proposed protocol in detail, including preliminary and protocol phases. Section \ref{sec5} demonstrates the security and privacy analysis, and Section \ref{sec6} evaluates the experimental results and performance. Finally, Section \ref{sec7} concludes the paper by summarizing the outcome and achievements of the proposed protocol.

\section{Literature Review}\label{sec2}

A lightweight and secure authentication scheme named Provably Secure and Lightweight Authentication Key Agreement (PSLA) is proposed by \citet{chai2023provably} based on the Shangyong Mima2 (SM2) Authenticated Key Exchange (AKE) Protocol stipulated by the State Cryptography Administration of China, which especially focuses on devices with limited computational capacity in the Smart Grid scenario. This scheme meets all the security needs and reduces computational overhead up to about 30\%-40\%,  but no threat scenario is discussed. 

\citet{Singh2023} propose a privacy-preserving multidimensional data aggregation and storage scheme (PP-MDA) using Paillier homomorphic encryption to aggregate encrypted multidimensional smart meter data. Communication and computation costs show a better result in this scheme in the performance analysis. They also proposed a secure and privacy-preserving data aggregation and classification (SP-DAC) model based on fog and cloud architecture in \cite{Singhs2023} where, besides HE, machine learning algorithms are also employed to classify smart home appliances for managing energy distribution. 

A Security Enhanced Lightweight Authentication and Key Agreement Framework (SE-LAKAF) for smart grid network based on the fuzzy extractor, elliptic curve cryptography (ECC), and AES-based hybrid cryptography is proposed by \citet{Mehta2023} which has optimal communication and computation overheads with improved security. Moreover, the proposed framework preserves user anonymity and resists well-known attacks such as impersonation, Man-in-the-Middle, and Denial-of-Service attacks. 

\citet{Xu2023} developed a privacy-preserving framework for smart metering systems using HE. Their proposed work utilizes different trust boundaries to analyze HE conﬁgurations for various scenarios in practical applications and also evaluates the feasibility by real-world implementation, which shows satisfactory results in terms of cost-effectiveness.

The protocol named EPri-MDAS is introduced by \citet{zhang2024epri} based on the ElGamal homomorphic cryptosystem and the GIFT-COFB block cipher to ensure data integrity and privacy without relying on a trusted authority (TA). It incorporates multidimensional data aggregation and secret sharing between smart meters and other entities to facilitate fault tolerance and secure data transmission.

The RPMDA scheme was proposed by \citet{liu2024rpmda} that uses the Chinese remainder theorem for efficient multidimensional data processing, a double-masking method for secure data aggregation, and a conditional anonymous certificateless signature algorithm for smart meter authentication. It offers superior computational efficiency and robustness, safeguarding data privacy.

\citet{zhang2024pfdam} proposed a {pri\-va\-cy-pre\-serv\-ing} Fine-grained Data Aggregation Scheme Supporting Multi-functionality (PDFAM) that combines Hash-based Message Authentication Code (HMAC) and public-key encryption for privacy-preserving data aggregation. The scheme is designed to ensure data integrity, unlinkability, and fault tolerance while resisting malicious data mining attacks.

\begin{table}[ht]
\centering
\caption{Outcomes and limitations of the related works}
\label{tab}
\begin{tabular}{@{}L{0.2\textwidth}
                   L{0.2\textwidth}
                   L{0.3\textwidth}
                    L{0.3\textwidth}@{}}
\toprule
 References & Methodology & Achievements & Shortcomings \\
\midrule
 PSLA \cite{chai2023provably}  & Shangyong Mima2 and Authenticated Key Exchange & Resilient to common attacks & Does not improve computing time and energy consumption\\
       PP-MDA \cite{Singh2023} & Paillier Cryptography & Supports multidimensional data aggregation & Does not consider possible attacks\\
        SE-LAKAF \cite{Mehta2023} & ECC-AES hybrid encryption & Resist identified attacks such as session key secrecy, DoS & Does no provide mutual authentication \\
        \citet{Xu2023}  &  Homomorphic encryption & Protects customer privacy under varying trust boundary & High computational overhead\\
        EPri-MDAS \cite{zhang2024epri} &  ElGamal cryptography & Resilient to internal and external
attacks & High computational complexity \\
       RPMDA \cite{liu2024rpmda} & Chinese remainder theorem & Stable system even when SMs malfunction &  No detailed analysis on the aggregated data\\
       PDFAM \cite{zhang2024pfdam} & HMAC and PKI & Resistant to malicious data mining attack & Substantial computational overhead\\
       P3HF \cite{kabir2024privacy} & Reversible watermarking and Paillier cryptography & Significantly low computational overhead & May be vulnerable to DoS attacks \\ 
\bottomrule 

\end{tabular}
\end{table}
A privacy-preserving protocol for high-frequency smart meters was introduced by \citet{kabir2024privacy} using difference expansion-based reversible watermarking and Paillier homomorphic encryption. It introduced an external entity called Encryption Server to perform HE to reduce the computation load of resource-limited smart meters. The acquired results show that the proposed scheme ensures security and user data privacy while consuming a low amount of energy and time for execution.

We present a concise overview of the aforementioned research articles in Table \ref{tab}, outlining their approach, noteworthy achievements, and recognized shortcomings.

\section{Overview of the System Model}\label{sec3}
In this section, we provide a system model to more clearly illustrate the roles and information interactions of each entity in our proposed protocol as well as their security and threat requirements.

\subsection{System Architecture}\label{sec3.1}
In the proposed protocol, the system model consists of three major entities: Smart Meter (SM), Data Aggregator (DA), and Control Center (CC). This system does not involve any trusted third party or external entity. The major roles of the three entities are described below:

\begin{itemize}
    \item 
     SMs are the core component of the system that is responsible for collecting fine-grained energy consumption data in real-time from each smart home using a home area network (HAN) \cite{Rind2023}. In this protocol, SMs are assumed to be fully trusted and tamper-proof.
    \item 
     DA is the middle entity or gateway that collects data from all the SMs located in that particular area, using a neighboring area network (NAN), and aggregates them before sending them to the final service center. It can be semi-honest or dishonest.
    \item  CC is the final destination of the aggregated data, where the aggregated data coming from the DAs through the wide area network (WAN) is analyzed and used for several purposes. Like DA, it can also be semi-honest or dishonest.
\end{itemize}

The whole system is designed  as a three-layer architecture which is shown in Fig. \ref{fig1}:
\begin{figure}
    \centering
    \includegraphics[width=0.8\textwidth]{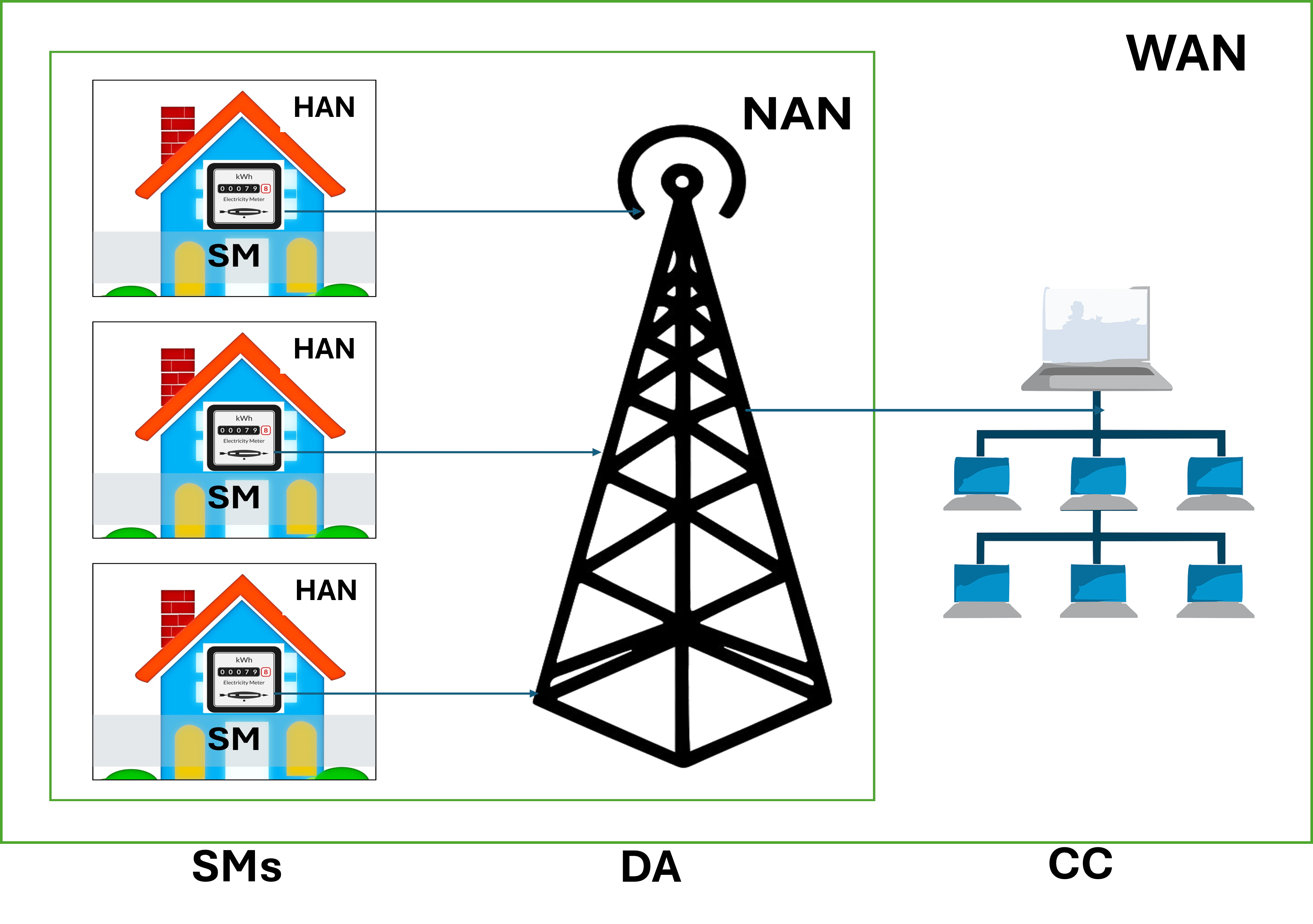}
    \caption{Smart Metering System}
    \label{fig1}
\end{figure}

\subsection{Security and Privacy Requirements}\label{sec3.2}
The security and privacy requirements for IoT-enabled SMs are what the SM needs to keep the system safe and private \cite{akkad2023information}. 
A privacy-preserved smart metering system must ensure confidentiality, integrity, and availability (CIA) that are vital for energy management, operation, protection, and communication \cite{hasan2023review}. 

The proposed protocol aims to achieve the following security and privacy requirements: 

 \begin{itemize}
     \item To limit access to transmitted data to only authorized parties, the system must guarantee data confidentiality. No external or internal adversaries should be able to reveal the contents; thus, the confidentiality of the data should be preserved.
     
     \item The protocol should guarantee data integrity by ensuring that the SM data is not manipulated or altered during transmission and storage. The data sent from the SM and received by the CC should be the same.
   
     \item The protocol should ensure that the energy consumption data received from the SM is authentic and from a legitimate source and that the DA can verify the identity of the SM and CC can verify the legitimacy of the DA. Authenticity verification is important to protect the system from unauthorized access.
 \end{itemize}

\subsection{Attack Model and Security Assumptions}\label{sec3.3}

To demonstrate the security and reliability of the proposed protocol, We considered well-known threats faced by smart metering systems in the real world. The attacker $\mathcal{A}$ in our protocol has the following capabilities:

\begin{itemize}
    \item SMs are considered fully trusted and temper-proof, while DA and CC are considered honest but curious or semi-trusted.
    \item The communication or data transmission during the initialization phase is assumed to occur in a secure channel, while the protocol phases take place over an open and vulnerable communication channel.
    \item $\mathcal{A}$ can obtain, eavesdrop on, modify, delete, or resend messages over the public channel, conducting cyber attacks such as Man-in-the-Middle (MitM), replay, impersonation, or Denial-of-Service (DoS) attacks.
    \item $\mathcal{A}$ can also conduct physical attacks like false data injection (FDI) during data transmission to gain access to the fine-grained SM data. 
\end{itemize}

\section{Proposed protocol}\label{sec4}

This section explores the fundamental components of the proposed protocol, followed by an in-depth description of both versions of the protocol, elaborating its design, algorithms, and processes phase by phase.

\subsection{Preliminaries}\label{sec4.1}
In the preliminary sections, we introduce reversible watermarking and AES cryptography as the key concepts and techniques that form the base of our proposed protocol.

\subsubsection{Reversible watermarking:}\label{sec4.1.1}
Reversible watermarking aims to send secret data over the communication channel in a secure manner. This is accomplished by embedding the watermark that can be accurately recovered at the receiving end \cite{kumar2022reversible}. A variety of hash functions, including SHA-512, SHA-256, and SHA-224, can be employed to generate the hash values required for computing the watermark.

\begin{itemize}
    \item Reversible watermarking using LSB-shifting: RLS is a technique that embeds watermark bits into the digital data utilizing LSB manipulation or insertion \cite{tran2022lsb}. That can be done by embedding one bit of watermark into the least significant bit (LSB) of the original data after shifting one bit to the left. 
    \item Reversible watermarking using difference expansion: The RDE approach is employed utilizing the redundancy of the data. The watermark bit is embedded into the difference value of two consecutive data points. The watermarked data is generated by utilizing the average and difference values between adjacent data \cite{kumar2022reversible}.
\end{itemize}

Both of these watermarking techniques are fully reversible and they follow three main steps: watermark generation, watermark embedding, and watermark extraction. 

\subsubsection{AES cryptography:}\label{sec4.1.2}
AES is a symmetric key cryptographic algorithm and a block cipher where the same key is used for encryption and decryption and it is known to both the sender and the receiver. The AES key can vary among 128, 192, and 256 bits \cite{gonzalez2023security} which takes 10, 12, and 14 rounds, respectively, and the fixed block size is 128 bits. The key data blocks and inputs are managed as arrays. Before generating the ciphertext, each array element is referred to as a state. Each round consists of 4 steps: SubBytes, ShiftRows, MixColumns and Add Round Key. The last stage of each round will not involve the MixColumns stage; instead, the remaining three stages will be repeated. The decryption process is the encryption process done in reverse \cite{andriani2018comparision}. 

\subsection{Protocol Phases}\label{sec4.2}
The proposed protocol has two different versions for low and {high-fre\-quen\-cy} smart meters which employ reversible watermarking based on LSB-shifting and difference expansion, respectively. In addition, both use AES cryptography for additional security. Both approaches consist of 3 phases that occur in the 3 entities of the system model depicted in Fig. \ref{fig1}.

An initialization phase is involved at the beginning of the procedure to generate the watermark and the keys needed for the encryption process. Additionally, we use pseudorandom numbers at every stage, derived from particular seeds for every stage.
\subsubsection{Initialization phase:}
We assume that there are $n$ SMs throughout the residential area.   The initialization phase performs the following tasks: 

\begin{itemize}
   \item Registration: The SMs must complete a registration process with CC before offering household electricity management services. In this procedure, the SM obtains a distinct identity, $\mathrm{SM}_i$. The number of SMs, $n$ must be odd for this protocol in order to make the watermark verification process work accurately (as explained in Section \ref{cc}). If $n$ is even, CC will add a dummy smart meter $SM_{n+1}$ during registration, which will always have 0 energy consumption.
   
   The DA initiates its registration with the CC before having the responsibility of managing all the SMs within a designated region.
   
    \item Data collection: When considering a total of $n$ SMs and $m$ time frames, each $\mathrm{SM}_i$ sends its energy consumption data, $d_{i,j}$, at each time period, $t_j$, where $i = 1,2,3,...,n.$ and $j=1,2,3,...m$. For each time frame, $t_j$, the total consumption of the residential area is $D_j = d_{1,j}+d_{2,j}+d_{3,j}+....+d_{n,j}$.

    \item Key generation and sharing: Each $\mathrm{SM}_i$ initializes its system and prepares to receive the key $K_w$ generated by CC for the SHA-2 cryptographic hash function, which will be used for producing the hash values involved in the watermark generation. For each time frame, the key $K_w$ and its time-stamp $t_j$ will be input into a one-way hash function. The key $K_w$ is shared with all the fully trusted SMs during this registration process.

    For AES encryption, each $\mathrm{SM}_i$ generates a symmetric key, $\mathrm{key}_{\mathrm{AES}}$, which is used for both encryption and decryption. It is shared with DA during the registration phase, allowing both entities to encrypt and decrypt data using the same key. There is no requirement for per-frame key distribution because the key is safely exchanged upon registration and is constant throughout all the time frames. This shared key simplifies key management while enabling mutual data access. 
     
    \item Watermark generation: The watermark $W$ is generated based on a Secure Hash Algorithm (e.g.\ SHA- 224, SHA-256, ore SHA-512). This cryptographic hash function uses $K_w$ and the time $t_j$ to generate a hash value $H_j=\mathrm{Hash}(K_w,t_j)$ for each time frame. Therefore, for $m$ time frames, there will be $m$ hashes. $K_w$ is available to all SMs and the CC to produce the necessary hash values. Finally, to generate the watermark, all the hash values are combined using the XOR operation: $W=H_1 \oplus H_2 \oplus \dots \oplus H_m$.

    This watermark ($W$) is a binary number of $m$ bits, and each bit, denoted as $w_j$, is used for the respective time frame $t_j$.
    
    \item Pseudorandom number generation: To add an extra layer of security, each SM data is encrypted with some pseudorandom numbers at each phase as an authenticator and privacy protector using a pseudorandom number generator (PRNG). A single seed sets the initial internal state of the PRNG. At each time frame $t_j$, the PRNG produces a new pseudorandom number and the internal state of the generator is also updated. Each $\mathrm{SM}_i$ generates three pseudorandom numbers at each time frame (or every other time frame) using three separate PRNG.
    
    In low-frequency SMs, the first pseudorandom numbers are obtained at $j = 1$, and, at every timeframe ($t_j$),  the pseudorandom generators are called to obtain the next pseudorandom numbers.
    
    Firstly, $n$ pseudorandom numbers $R3_{i,j}$ are generated for each $\mathrm{SM}_i$, which are specific for each time frame $t_j$. The value $R3_{i,j}$ is only known by $\mathrm{SM}_i$ and CC, and DA has no access to this. Similarly, another set of $n$ pseudorandom numbers $R1_{i,j}$ is generated for each $\mathrm{SM}_i$, which is specific for the time frame $t_j$. $R1_{i,j}$ is only known to $\mathrm{SM}_i$ and DA, and CC has no access to this. Finally, another pseudorandom number $R2_{j}$ is generated for each time frame $t_j$, which is known only by DA and CC.

    In high-frequency SMs, the first pseudorandom values are generated at $j = 2$. At each even time frame $t_{2j}$, the PRNGs are called to obtain the next pseudorandom numbers.
    Two pseudorandom numbers denoted by $R3_{i,2j-1}$ and $R3_{i,2j}$ are generated for time $t_{2j-1}$ and $t_{2j}$ which are known by both SM and CC. For the other two pseudorandom number generators, only one value is required at each even time frame. $R1_{i,2j}$ is known by SM and DA and $R2_{2j}$ by DA and CC. 
\end{itemize}



\begin{figure}[ht]
    \centering
    \includegraphics[width=1\linewidth]{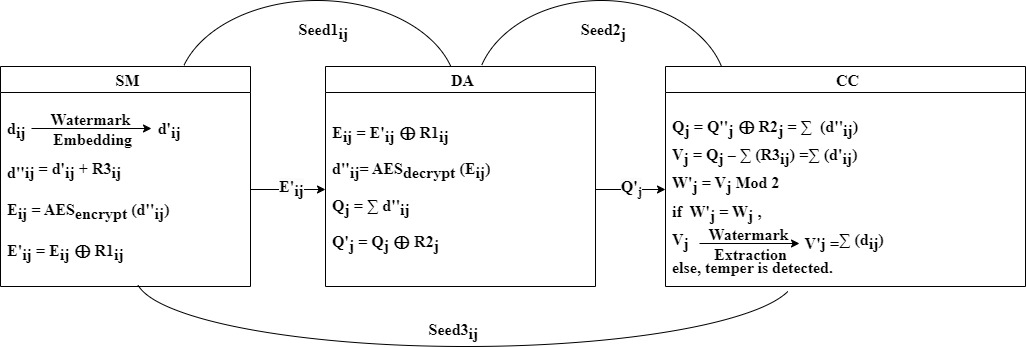}
    \caption{Protocol phases for low-frequency SMs}
    \label{fig2}
\end{figure}

\begin{figure}
    \centering
    \includegraphics[width=1\textwidth]{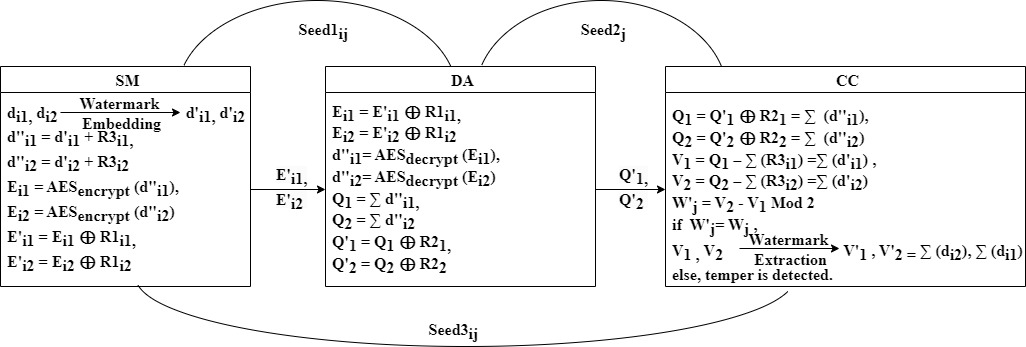}
    \caption{Protocol phases for high-frequency SMs}
    \label{fig3}
\end{figure}

The overall demonstration of the various steps of the proposed protocol for low-frequency and high-frequency SMs can be seen in Figure \ref{fig2} and \ref{fig3}.

\subsubsection {Smart meter phase:} 

Each $\mathrm{SM}_i$, at each time frame $t_j$, obtains $d_{i,j}$, which is the energy consumption data for that time frame. In the proposed algorithm, we convert any floating value of $d_{i,j}$ into an integer prior to watermark embedding (with a few fixed decimal positions). In case of high-frequency smart meters, since difference expansion is used, we also multiply the (integer) data values by 2, ensuring that all computed averages and differences remain integers, eliminating the need for rounding operations in subsequent calculations. 

For low-frequency SMs, the particular watermark bit $w_j$ for that time frame is embedded into the least significant bit of meter reading $d_{i,j}$. To do this, we just need to multiply $d_{i,j}$ by 2 and then add $w_j$ with it. Hence, the watermarked data is generated as $d'_{i,j}=2d_{i,j}+w_j$. After that, we add the pseudorandom number $R3_{i,j}$ (which is shared only between $\mathrm{SM}_i$ and CC) with $d'_{i,j}$ and obtain $d''_{i,j}$. 

In $\mathrm{SM}_i$, $d''_{i,j}$ is encrypted using the AES encryption algorithm. The first step is to generate fixed-size keys (128, 192, or 256 bits) for encryption. As AES uses symmetric key cryptography, both encryption and decryption will be accomplished with the same key, $\mathrm{key}_{\mathrm{AES}}$. In this encryption technique, the input data is divided into several blocks (128-bit in general) and then combined with the initial round key using a bitwise XOR operation in the primary round. After that, a series of rounds takes place as described in Section \ref{sec4.1.2}. On completion of the final round, the ciphertext block is generated, which is the encrypted form of the original plaintext block producing the encrypted data $E_{i,j}$. Before sending the encrypted watermarked data to the data aggregator, another pseudorandom number $R1_{i,j}$ generated by the PRNG is used to further encrypt $E_{i,j}$ using the XOR operation. The final output $E'_{i,j}$ is sent to DA.

In high-frequency smart meters, we divide the \( m \) time frames into \( \frac{m}{2} \) pairs, such that each iteration considers only two consecutive time frames. For each time frame, the relevant data points for each \( SM_i \) are \( d_{i, 2j-1} \) and \( d_{i, 2j} \), where \( j = 1, \dots, \frac{m}{2} \). To embed the watermark bit, we apply the difference expansion technique (a generalized integer transform) explained in Algorithm \ref{alg2}. At each even time frame \( 2j \), one bit of the watermark is embedded in the least significant bit of the difference between \( {d}_{i, 2j-1} \) and \( {d}_{i, 2j} \). This operation produces the watermarked data \( d'_{i, 2j-1} \) and \( d'_{i, 2j} \) via the difference expansion transform. After that, we add $R3_{i, 2j-1}$ and $R3_{i, 2j}$ to $d'_{i, 2j-1}$ and $d'_{i, 2j}$ respectively. After AES encryption, we obtain $E_{i,2j-1}$ and $E_{i,2j}$ and, after performing XOR with $R1_{i, 2j}$, the values $E'_{i,2j-1}$ and $E'_{i,2j}$ to be transmitted are obtained.

\begin{algorithm}[ht]
\caption{RDE Embedding}
\label{alg2}
\textbf{Input:} $w_{2j}$, [${d}_{i, 2j-1}$], [${d}_{i, 2j}]$ \\
$\mathrm{avg} \gets \frac{2({d}_{i, 2j-1} + {d}_{i, 2j})}{2}$\\
$\mathrm{diff} \gets 2( {d}_{i, 2j-1} - {d}_{i, 2j})$\\
$d'_{i, 2j-1} \gets \mathrm{avg} - \frac{\mathrm{diff}}{2}$ \\
$d'_{i, 2j} \gets \mathrm{avg} + \frac{\mathrm{diff}}{2} + w_{2j}$ \\
\textbf{Output:} [$d'_{2j-1}] , [d'_{2j}]$ 
\end{algorithm}

\subsubsection {Data aggregator phase:} 
 
DA receives the watermarked encrypted data $E'_{i,j}$ from all $n$ low-frequency $\mathrm{SM}_i$ and performs the XOR using $R1_{i,j}$ to obtain $E_{i,j}$. After that, AES decryption takes place, and the watermarked data $d''_{i,j}$, which has the pseudorandom number $R3_{i,j}$ added to it, is obtained. Then, DA aggregates all the $d''_{i,j}$ coming from different $\mathrm{SM}_i$ together using summation and obtains $Q_j$ = $\sum d''_{i,j}$. Next, DA encrypts it again using another pseudorandom number $R2_{j}$, which is known only to DA and CC. The obtained output $Q'_j$ is finally transferred to the control center. 

The same procedures are followed for high-frequency smart meters, resulting in $Q'_{2j-1}$ and $Q'_{2j}$ by performing XOR with $R2_{2j}$.

\subsubsection {Control center phase:} 
\label{cc}
Upon receiving $Q'_j$, CC obtains $Q_j$ = $\sum d''_{i,j}$ performing XOR decryption with $R2_{j}$. After that, CC subtracts $\sum R3_{i,j}$ and obtains $V_j = \sum d'_{i,j}$ which is the sum of the watermarked data. At this point, it is required to perform a watermark verification process by recalculating the watermark bit $w_j$ using the same key for the hash function which is shared with all the smart meters. For low-frequency SMs, we verify it by comparing $\sum w_j$ with ($V_j\mod2$). If they are not the same, that indicates the tampering of the data and is immediately declined. If they are equal, we proceed with the watermark extraction and obtain the original summation of the data $D_j$ = $\sum d_{i,j}$ = $V_j$ - $\sum w_j$. 

For high-frequency SMs, $V_{2j}$ and $V_{2j-1}$ are obtained after subtracting $R2_{i, 2j}$ and then subtracting $\sum R3_{i,2j-1}$ and $\sum R3_{i,2j}$ respectively from $Q_{2j}$ and $Q_{2j-1}$, respectively. The watermark is then validated by comparing the LSB of $V_{2j} - V_{2j-1}$ with $w_{2j}$. If they are not equal, it is assumed that the data was tampered with and is declined immediately. 

If the LSB is correct, CC extracts the watermark using the inverse transform as shown in Algorithm \ref{alg6} and computes the sum of the plaintext data $D_{2j-1}$ = $\sum {d}_{i, 2j-1}$ and $D_{2j}$ = $\sum {d}_{i, 2j}$ for the time frame $t_{2j-1}$ and $t_{2j}$. This process is repeated $m/2$ times to obtain all aggregated energy consumption data of total $m$ time frames.

The watermark verification, both for the low-frequency and high-frequency cases, is true only if $n$ is odd. In $n$ was even, the relevant LSB would always be 0, making the verification process invalid.

\begin{algorithm}[ht]
\SetAlgoLined
\caption{RDE Watermark extraction}
\label{alg6}
\KwIn{$V_{2j-1},V_{2j}, w_{2j}$}
$w'_{2j}\gets \left(V_{2j} - V_{2j-1}\right) \mod 2$\;
\If{$w'_{2j} = w_{2j}$}{
  $\mathrm{diff} \gets \displaystyle\frac{V'_{2j} - V'_{2j-1}-n (w_{2j})}{2}$\;
  $\displaystyle{\mathrm{avg} \gets \frac{V'_{2j}+V'_{2j-1}-n(w_{2j})}{2}}$\;
  $\widetilde D_{2j-1} \gets \mathrm{avg} - \displaystyle  \frac{\mathrm {diff}}{2}$\;
  $D_{2j-1} \gets \displaystyle\frac{\widetilde D_{2j-1}}{2}$\;
  
  $\widetilde D_{2j} \gets \mathrm{avg} + \displaystyle \frac{\mathrm {diff}}{2}$\;
  $D_{2j} \gets \displaystyle\frac{\widetilde D_{2j}}{2}$\;
  \Return $D_{2j-1}, D_{2j}$\;
}
\Else{
  \Return Error (tampering detected!)\;
}
\end{algorithm}

\section{Security and Privacy Analysis} \label{sec5}
In this section, a security and privacy analysis is conducted for the proposed protocol in terms of the requirements: confidentiality, integrity, and availability (CIA), as well as resistance to the most popular attacks.

In the proposed protocol, the SM embeds a watermark on the real, fine-grained data and later encrypts it twice, first using AES encryption and then with a pseudorandom number to maintain proper confidentiality. The DA does not have access to the $R3_{i,j}$, therefore, it cannot retrieve the original or watermarked data even after AES decryption. CC gets only aggregated value for each residential area after extracting the watermark successfully and it can never get the fine-grained user data for each SMs. Therefore, data confidentiality remains ensured.

During each communication phase, specific pseudorandom numbers are used, computed with the PRNG that were initialized using the seeds shared among entities during the initialization pahse. At each time frame, for each phase, these pseudorandom values are used; if they do not match, data will be corrupted upon decryption, thus to guaranteeing authenticity. The seeds are different for all SMs, which acts as an authentication parameter of each SM to make it easy to verify the authenticity of data at every phase. Reversible watermarking has been used to ensure end-to-end data integrity and can successfully recover the accurate original energy consumption data. 

The data goes through several levels of security and privacy protection, including watermarking, pseudorandom number addition, and AES encryption, so that the original data cannot be manipulated by the adversary, not even a compromised DA, or CC. Since data is encrypted, an eavesdropper would never access the original data. The pseudorandom numbers that are added in each time frame are different. Even with the same data, different encrypted results will be achieved.

The verification of timestamp (in the watermark bit and also by updating the different pseudorandom numbers) prevents replay attacks. Since the pseudorandom numbers change at every timestamp, the replayed data will be wrong, and the decrypted data will typically have a forbidden value. 

Man-in-the-middle attacks can be detectable and avoided by preventing the attacker's access to the private keys for watermarking, encryption, and seed generation. If a MitM attack takes place in DA, it cannot achieve any of its goals, because the relevant pseudorandom value would not match. If it attacks CC, the attacker will still get only the aggregated value, from which it is impossible to gain access to the fine-grained SM data. 

Both versions of the protocol limit the ability to decrypt and manipulate data by giving each component specified access privileges. Even if they are successful in impersonating one entity, they will not be able to access or decode various levels of data. Timestamp verification ensures that only current data is accepted, which means if attacks flood the system with outdated data, they will be most probably rejected based on the timestamps. Furthermore, the proposed protocol eliminates external entities such as trusted authority in \cite{zhang2024pfdam,liu2024rpmda} or encryption server in \cite{kabir2024privacy}, which significantly limits the probability of a DoS attack by reducing the attack surface.

Table \ref{tab3} presents a comparison of recent studies with the proposed protocol in terms of their resistance to attacks, demonstrating that the proposed protocol effectively prevents most potential attacks compared to other protocols.

\begin{table}
    \centering
    \begin{tabular}{ccccccc}
    \hline
        Protocols & Eavesdropping & Replay & MiTM  & Impersonation & FDI & Reduced DoS \\ \hline
         EPri-MDAS \cite{zhang2024epri} & \Checkmark & \Checkmark & \Checkmark & \XSolid  & \Checkmark & \XSolid \\
         RPMDA \cite{liu2024rpmda}& \Checkmark & \Checkmark & \Checkmark & \Checkmark  & \XSolid & \XSolid \\
         PFDAM \cite{zhang2024pfdam}  & \Checkmark & \Checkmark & \Checkmark & \XSolid &\Checkmark & \XSolid \\
         P3HF \cite{kabir2024privacy} & \Checkmark  & \Checkmark & \Checkmark & \Checkmark &\Checkmark & \XSolid \\
         Proposed protocol &\Checkmark  & \Checkmark & \Checkmark & \Checkmark & \Checkmark  & \Checkmark\\ \hline
    \end{tabular}
    \caption{Comparison in terms of attack resistance}
    \label{tab3}
\end{table}

\section{Experiments and Results}\label{sec6}

We have carried out different experiments focused on assessing the proposed protocol for low-frequency and high-frequency smart meters that generate data in a specific time interval. We have performed our experiments for varying numbers of SMs (100 to 1000) to calculate the average computational cost at each SM and compare it with other recent protocols.
  
\subsection{Experimental Setup}
\begin{figure}
    \centering
    \includegraphics[width=0.5\linewidth, angle=90]{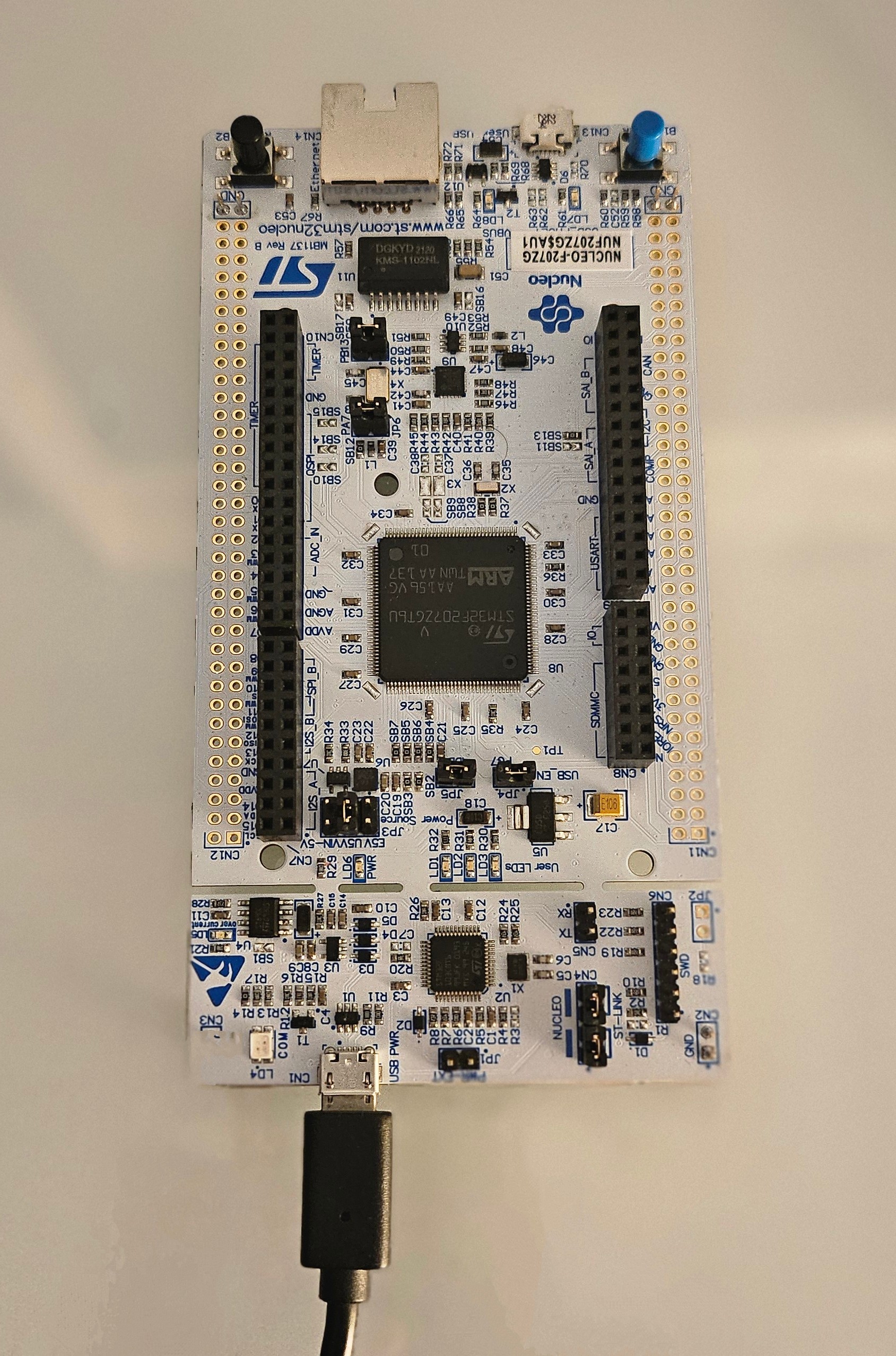}
    \caption{The Nucleo-f207zg microcontroller board used for simulation}
    \label{fig4}
\end{figure}

The suggested privacy-preserving protocol has been executed and assessed leveraging real low-frequency and high-frequency SM data in a simulated scenario. Using the RIOT operating system on the Nucleo-f207zg development board, we experimented with smart meters in a home environment to evaluate the protocol's effectiveness and performance, as shown in Figure \ref{fig4}.

The Nucleo board performed as an SM that collects energy consumption data every 1 hour in low-frequency SM and 60 seconds in high-frequency SM. The data was first watermarked using the two different reversible watermarking techniques for two versions of the protocol and then encrypted using AES encryption to ensure data protection.

\subsection{Performance Evaluation}

The performance of the protocol was evaluated by measuring the time required for the initialization phase, which includes watermark generation and key generations. We also measured the time taken by the SM at each iteration of phase 1.

The table presents insights regarding the computational efficiency of different configurations of SHA (224, 256, and 512 bits) and AES (128, 192, 256 bits) in the initialization and iteration phases of the protocol. Results were averaged over 1000 iterations for accuracy.

\begin{table}[ht]
    \centering
    \caption{Initialization and iteration times}
    \begin{tabular}{@{}ccc@{}}
        \toprule
        \multicolumn{3}{c}{\textbf{Initialization Time}} \\ \midrule
        \textbf{SHA} & \textbf{Time for both RLS and RDE (ms)} & \\ \midrule
        224 & 19.29 & \\
        256 & 19.40 & \\
        512 & 74.58 & \\ \midrule
        \multicolumn{3}{c}{\textbf{Iteration time for 1 SM}} \\ \midrule
        \textbf{AES} & \textbf{RLS (ms)} & \textbf{RDE (ms)} \\ \midrule
        128 & 0.068 & 0.074 \\
        192 & 0.075 & 0.079 \\
        256 & 0.080 & 0.082 \\ 
        \bottomrule
    \end{tabular}
    \label{tab:initialization_iteration_times}
\end{table}
The initialization time appears to correlate with the complexity and bit length of the SHA algorithm used. The initialization times of SHA-224 and SHA-256 are comparable (19.29 ms and 19.40 ms, respectively), suggesting that they are computationally effective and suitable for applications requiring fast initialization. However, SHA-512's initialization time (74.58 ms) is noticeably higher, reflecting the greater computational load that the greater hash size brings. On the other hand, the iteration time is determined by AES encryption standards (AES-128, AES-192, and AES-256), which varies slightly between reversible watermarking techniques based on RLS and RDE. RLS takes 0.068 ms for each iteration for AES-128, however, RDE takes 0.074 ms, which is a little longer. 
\begin{figure}
    \centering
    \includegraphics[width=1\linewidth]{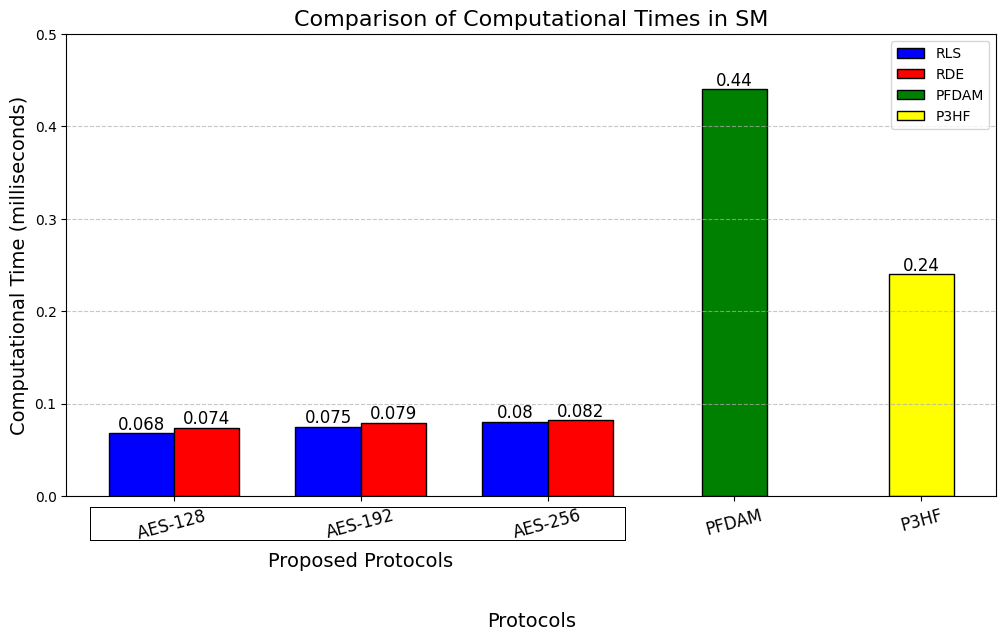}
    \caption{Comparison Graph}
    \label{fig5}
\end{figure}

As the key length grows, AES-192 and AES-256 show increased iteration times; AES-256 takes the longest (with RLS taking 0.080 ms and RDE taking 0.082 ms).  This trend indicates that the longer AES key lengths need more processing time. 

The bar chart in Figure \ref{fig5} presents a comparative analysis of the computational times (in milliseconds) for two cryptographic protocols P3HF \cite{kabir2024privacy} and PFDAM \cite{zhang2024pfdam}, as well as the two proposed versions of our protocol, which shows that the computational time in all the AES versions, and both for RLS and RDE watermarking, the proposed protocol has significantly lower computational cost than the other two protocols. 

The initialization time may seem longer, but in the proposed protocol, it is a fixed cost that can be distributed over multiple iterations or sessions. Once the protocol is initialized, subsequent operations benefit from this initial setup without incurring the same overhead, and the time for subsequent operations is extensively low. This makes both versions of our protocol the most efficient and cost-effective approach toward privacy-preserving data aggregation in smart metering systems without compromising security and privacy.

In addition, the reduced communication overhead is another key advantage of the proposed protocol compared to other protocols and the plaintext equivalents. Unlike other encryption schemes that may increase the payload size, AES maintains a fixed data size because it operates as a block cipher. The pseudorandom number addition and reversible watermarking are embedded at the algorithmic and cryptographic level without altering the total data size. Hence, the number of packets transmitted over the network remains the same as in plaintext communication, resulting in zero communication overhead.

\subsection{Limitations and Future Directions}
Although the number of packets transmitted in our communication system is similar to that found in plaintext scenarios, we have observed that the overall communication cost, when measured in bits, shows potential for significant improvement relative to recent protocols. Our future research directions will focus on incorporating more lightweight and modern cryptographic methods, specifically exploring the use of ASCON cryptography as an alternative to traditional cryptographic algorithms. This future work will aim to enhance the efficiency of the protocol by reducing the communication overhead while maintaining the overall efficacy of the system.

\section{Conclusion}\label{sec7}
This paper proposes and simulates a robust privacy-preserving protocol for smart meters in the RIOT operating system, focusing on data confidentiality, integrity, and authenticity. By applying a joint technique combining reversible watermarking and AES cryptography, the proposed protocol ensures that sensitive fine-grained meter data remains inaccessible to unauthorized attackers. The protocol provides two versions (RLS and RDE) for the low- and high-frequency smart meters that use reversible watermarking using LSB-shifting and different expansion, respectively.

We have demonstrated the protocol's computational efficiency through systematic simulations using the Nucleo microcontroller in the RIOT OS for different versions of the SHA hash function and AES key size. Our findings highlight the protocol's practicality for real-world applications, as it not only stands out in terms of computational cost but also successfully implements robust security measures to defend against a variety of potential threats. The results evidently indicate that the proposed protocol surpasses existing methods, making it a better choice for privacy-preserving data aggregation in smart metering systems.

\section*{Acknowledgments}
The authors acknowledge the funding obtained by the grants: Detection of fake newS on SocIal MedIa pLAtfoRms (DISSIMILAR) from the EIG CONCERT-Japan (PCI2020-120689-2, Government of Spain), the ``SECURING'' project (PID2021-125962OB-C31) funded by the Ministerio de Ciencia e Innovación, la Agencia Estatal de Investigación and the European Regional Development Fund (ERDF), as well as the ARTEMISA International Chair of Cybersecurity (C057/23) and the DANGER Strategic Project of Cybersecurity (C062/23), both funded by the Spanish National Institute of Cybersecurity through the European Union – NextGenerationEU and the Recovery, Transformation and Resilience Plan.

\renewcommand{\bibsection}{\section*{\refname}}  

\bibliographystyle{unsrtnat}  

\end{document}